\documentstyle[12pt,epsfig]{article} \begin{document}
\begin{center} 
{\bf Glassy behaviour in short range lattice models without quenched disorder} 
\vskip 1cm 
M\'ario~J.~de~Oliveira.${}^\dagger$ 
and Alberto~Petri ${}^\ddagger$\footnote{Author to whom correspondence 
should be addressed; e-mail: petri@idac.rm.cnr.it}\\
${}^\dagger$ Instituto de F\'{\i}sica, Universidade de S\~ao Paulo, 
Caixa Postal 66318,
05315-970 S\~{a}o Paulo SP, Brazil

${}^\ddagger$ Consiglio Nazionale delle Ricerche, Istituto 
``O.M.~Corbino'',  Via del Fosso
del Cavaliere 100, 00133 Roma, Italy \\and\\ Dipartimento di Fisica and Istituto
Nazionale per la  Fisica della Materia, Universit\`a ``La Sapienza'', P.le
A.~Moro 2, 00189 Roma, Italy\\ 
\end{center} 
\begin{center} 
{\bf  Abstract} 
\end{center} 
\par 
We investigate the quenching process in lattice systems with short range
interaction and several crystalline states as ground states.
We consider in particular the following systems on square lattice:\\
- hard particle (exclusion) model\\
- $q$ states planar Potts model.  \\
The system is initially in a homogeneous disordered phase
and relaxes toward a new equilibrium state
as soon as the temperature is rapidly lowered. 
The time evolution can be described numerically by a stochastic process such 
as the Metropolis algorithm.  
The number of pure, equivalent,
ground states
is $q$ for the Potts model and $r$ for the hard particle model, and 
it is known that for $r$, or $q \ge d+1$, the final equilibrium state may be
polycrystalline, i.e. not made of a uniform phase. We find that 
in addition $n_g$ and $q_g$ exist such that for $r > r_g$ or $q > q_g$ 
the system
evolves toward a glassy state, i.e. a state in which the ratio of the
interaction energy among the different crystalline phases to the total
energy of the system never vanishes; moreover we find indications 
that $r_g=q_g$. We infer that 
$q=q_g$ (and $r=r_g$) corresponds to the crossing from second order
to discontinuous transition in the phase diagram of the system.
\newpage
\begin{center}
{\bf  Introduction}
\end{center}
A large variety of materials occurs as glass or amorphous (Elliot 1990)
but mechanisms driving matter into such kind of states 
are far from clear. Because of the inherent complexity of the problem, 
much effort has been devoted also to the investigation of simple models
like lattice 
models. These systems may reproduce at least a few of the characteristic features of 
glassy behaviour,
however they are generally characterized by quenched disorder or
competing interactions. 
In contrast with real glass former such model systems are not able to 
crystallise and therefore cannot  give any
information on possible transitions from glassy to crystalline behaviour.

In this work we show that also simple lattice models with short range 
interaction and no quenched disorder may enter a kind of glassy phase.   
In particolar we show in section $2$ that if 
a $q$ state Potts system is quenched from infinite to
zero temperature, it relaxes toward a state of minimum energy  
only if $q$ is small enough. Otherwise it slowly equilibrates into a state 
with energy well above the ground state. Analogous observations
on different lattice models are elaborated in section $3$ and indicate 
that such behaviour may be universal.

\begin{center} 
{\bf  The Potts model}
\end{center}

Consider a regular lattice in which each site can stay in one of $q$
distinct states, or colours. To each site one may associate a variable $\eta_i$
that takes the values $0,1,2,\dots,q-1$. The energy of a configuration is
given by (Potts 1952)
\begin{equation} 
E=-\epsilon \sum_{(ij)}(\delta_{\eta_i,\eta_j}-1),
\end{equation}
where the
summation is over pairs of nearest neighbour sites. If two nearest
neighbour sites have the same colour their contribution to the total
energy vanishes, otherwise they contribute with an energy $\epsilon$.
There are therefore $q$ distinct ground states; in each of them all sites in the lattice
have the same colour and the total energy takes the absolute
minimum value,  $E=0$. These states are crystalline states. 

Let
$n_p$ the number of site pairs such that the two sites have distinct
colours, that is
\begin{equation}
n_p= \sum_{(ij)}(1-\delta_{\eta_i,\eta_j}),
\end{equation}
thus
\begin{equation}
E=\epsilon \ n_p
\end{equation}
and in a crystalline state $n_p=0$.
A polycrystalline state is composed of clusters of crystalline states of
distinct colours separated by $n_p$ site pairs of different colours. 
The number $n_p$ however must be small. More precisely
we call a state polycrystalline when $\nu=n_p/N \rightarrow 0$ as the number of lattice sites $N
\rightarrow \infty$, so that the total energy per site $u=E/N \rightarrow 0$,
the same  of the crystalline state.

We have simulated a quench to zero temperature as follows (Metropolis {\it et al.} 1953). 
At each step of simulation a site is chosen at random and its new colour is chosen at random. If the energy
does not increase, the new colour is taken as the new state. The initial
configuration of the system is a high temperature one, that
is each site has a colour chosen at random.

We have studied the time behaviour of the 
energy per site
$u=E/N$ after the quench on a square lattice for several values of $N=L^2$. 
It is well known that the dynamics of the Potts model usually gets pinned 
very soon into a non equilibrium state because of the finite 
size of  lattices used in the simulations (Glazier {\it et al}. 1990).
In order to escape this trap we have used both periodic boundary conditions
and fixed boundary conditions. In the latter, all 
sites on the lattice boundary
are frozen in a given state, say the $\eta=0$ state. By this way one can  
discriminate finite  size and boundary effects in the behaviour of the system. 
In fact we have assumed that, when plotting energy vs time, the superimposing 
section of curves resulting from different simulation conditions, may be 
identified with the genuine relaxation of the system. 
We have  
extrapolated the behaviour for $L\rightarrow \infty$ from the $1/\sqrt(t)$ part
of these curves.
As examples, the plots resulting from $q=3$ and $q=7$ are shown in figures~1~and~2. 

From simulations on a square lattice we have drawn the following conclusions:\\
a) for $q=2$  the system relaxes toward a crystalline 
state: $n_p \rightarrow 0$ and   
\begin{equation}
u \simeq t^{-1/2};
\end{equation}\\
b) for $q=3$ the state is polycrystalline: $n_p \neq 0$ but
$\nu=n_p/N\rightarrow 0$, 
\begin{equation}
u \simeq t^{-1/2}.
\end{equation}
This agrees with the Lifshitz proposition (Lifshitz 1962) that
the final states  are  non pure crystalline  when $q\ge d+1$ 
($q \ge 3$ in $d=2$).
However the proposition does not say anything about the
energy of these states;\\
c) for $q > 4$ we observe that $u \rightarrow u^*>0$ when $t \rightarrow
\infty$, since $\nu \rightarrow \nu^* > 0$, i.e.
\begin{equation}
u \simeq t^{-1/2} +u^*.
\end{equation}
The final state is not polycrystalline and, in addition, its energy is above the
energy of the ground state (in fact it is seen to be a monotonic function
of $q$). We call such pinned state a  glassy state.\\
d) for $q = 4$  the asymptotic behaviour 
is well fitted by 
\begin{equation} 
u \simeq t^{-1/2} \ln(t),
\end{equation} 
Preliminary results from simulations of quenches at $T>0$ indicate
that the observed features persist as far as $T$ is not too large.
\begin{center}
{\bf 
Hard core particles on a lattice}
\end{center}

Extension of the above observations to  
 different lattice models, also characterized by short range 
interaction and absence of quenched disorder, leads to remarkable conclusions.
We discuss here some results 
reported in the literature for a class of lattice gas models, namely exclusion 
models with random sequential adsorption and in-plane diffusional dynamics.

These models consist of a regular lattice in which particles are placed
at the sites of the lattice. The deposition of a particle in a site
blocks that site and the sites in a certain neighbourhood, whereas the energy is simply given by
\begin{equation} 
E=-\mu (n -n_{cp}) 
\end{equation} 
where $\mu>0$ is the chemical potential, $n$ is
the total number of deposited particles and $n_{cp}$ is the number of
deposited particles in the closest packed configuration. The configuration
of closest packing is a ground state configuration because the 
number of particles
gets maximum so that the energy $E$ is minimum ($\mu >0$) (in
analogy with the Potts model the ground state energy is arbitrarily set
to zero).

If a particle blocks the deposition of other particles in a certain
neighbourhood the ground state is not unique. When for instance
nearest neighbours sites are blocked on a square lattice (NN model) 
then
there are two ground states corresponding to place the particles on either
one  or the other of two sub-lattices. If a particle
blocks the nearest neighbours as well the next nearest neighbours (NNN
model) then there will be four relevant ground states, corresponding to four different
sub-lattices. If besides nearest neighbours and next nearest neighbours
the particle blocks also the next next nearest neighbours (NNNN model) then the
number of ground states will be five.

Once deposited, particles may diffuse along the lattice, with the prescription 
that they cannot overlap  other particles.  This process, called diffusional 
relaxation,  makes possible to depose more particles and, in principle, 
the system enter an ordered phase, coming eventually to the closest packed
state where density cannot further increase. 

Depending on the  model (NN, NNN, NNNN, etc.),  the final state may be 
crystalline,
polycrystalline or glassy, and can be classified in a way similar to 
the case
of the Potts model discussed above. Let $n_s$ be the number of sites at the 
surface
of clusters of sites in the same state, i.e.  occupying the same
sub-lattice:
\begin{equation}
n_s=n_{cp}-n.
\end{equation}
In analogy with the Potts model,
if $n_s=0$ the state is  monocrystalline. If $n_s\ne0$ but $n_s/N
\rightarrow 0$ as $N \rightarrow \infty$, the state is 
polycrystalline, and is glassy in the opposite case.

We briefly elaborate here on some results reported in 
the literature by Wang, Nielaba and Privman (1993, 1993a), and by
Eisenberg and Baram  (1998, 2000).
In the simulation a lattice initially empty is filled sequentially by
placing particles at the non excluded sites. Deposited particles may diffuse
to next neighbour sites if these are not blocked or occupied. 
Although the implementation of this dynamics varies slightly from one author 
to the other, their results are of
general validity.
Let us denote by $r$ the number of ground states so that $n_{cp}=N/r$,
and let $u=E/N$.
By analysing the papers cited above one concludes that:

a) for r=2 (NN model)  $ n_s \rightarrow 0$  and $u \rightarrow 0$
(Wang {\it et al.} 1993) 
 since
\begin{equation} 
u_s \simeq t^{-1/2};
\end{equation}

b) for $r=4$ (NNN model),  $n_s \ne 0$ but
$n_s/N \rightarrow 0$  as $N \rightarrow \infty$,  
(Wang {et al.} 1993a, Eisenberg and Baram 1998) and $u \rightarrow 0$ as
\begin{equation} 
u_s \simeq \ln(t)t^{-1/2};
\end{equation}

c) for $r=5$, $u \rightarrow u^*>0$ and 
\begin{equation} 
u_s  \approx t^{-1}+u^*_s;  \ \ u^*_s>0.
\end{equation}
This $t^{-1}$ behaviour has been observed in the very late stage 
(Eisenberg and Baram 2000),
when the (finite) system 
is almost completely filled, and therefore is not to be compared 
with the asymptotic, size independent, behaviour $t^{-1/2}$ of the Potts model.
The relevant point here is that $u \rightarrow u^* >0$ when 
$t \rightarrow \infty$, if $r>4$. Frozen states with density lower than in the 
closest packing have been observed recently also in different exclusion models
(Grigera {\it et al.} 1997, Fusco {\it et al.} 2001).
\begin{center}
{\bf  Discussion and conclusions}
\end{center}

The results obtained for the exclusion model are very similar to those
obtained for the Potts model.
An important point to stress is that the two models are predicted to have
the same symmetries 
and thus the same critical behaviour (Domany {\em et al.} 1977). 
However they are intrinsically different 
because there are features of the Potts model, like  surface tension and 
spontaneous magnetization,  that 
are not shared by the hard particle model.
Similarities in their behaviour support therefore the idea of $q_g=4$   
(and $r_g=4$)
as an upper bound for polycrystalline behaviour also in the Potts model.
 
From the above considerations it can be argued that these different behaviours 
are actually related to the equilibrium phase diagram of the system.
In fact, in two dimensions, for both Potts (Wu 1982) and hard particle models (Runnels 1983) 
the upper critical number of ground states is four.

Another important point to remark is that
the appropriate procedure to get the limiting energy 
is to take first the thermodynamic limit
$L\rightarrow \infty$, and then the infinite time limit $t \rightarrow
\infty$. 
This is a very important point that  makes possible the observation of 
glassy phases in the kind of systems considered here. In fact
systems  with only short range interaction and no quenched disorder 
cannot have metastable states in the usual sense (Griffiths 1964, Yeomans 1992).
  
In conclusion we have shown that  dynamics of simple 
lattice models  exhibits a transition from (poly) crystalline to glassy
behaviour when the number of ground states increases. 
Investigations on the Potts model and on exclusion models support the 
hypothesis that the critical number of states at which the transition 
takes place is related to the equilibrium  properties
and, for systems with this  symmetry,  is four.
This work has been supported by grants from CCInt-USP (Brazil) and CNR (Italy).

\begin{center}
{\bf References}
\end{center}
Domany  E., Schick M. and Walker J.S., 1977, {\it Phys. Rev. Lett.} 
{\bf 38}, 1148.\\
Eisenberg E. and Baram A., 1998, {\it Europhys. Lett.} {\bf44}, 168.\\
Eisenberg E. and Baram A., 2000, {\it J. Phys.} {\bf A 33}, 1729.\\
Elliot S.R., 1990, {\it Physics of amorphous materials} (New York: John Wiley \& Sons Inc.).\\
Fusco C., Gallo. P., Petri A. and Rovere M., 2001, {\it J. Chem. Phys.} 
{\bf 114}, 7563.\\
Glazier J.A., Anderson M.P. and Grest G.S., 1990, {\it Phil. Mag.} \\
{\bf B 62}, 615 and refs. therein.\\
Griffiths R.B., 1964, {\it J. Math. Phys.} {\bf 5}, 1215.\\
Grigera S.A., Grigera T.S. and Grigera J.R., 1997, {\it Phys. Lett.} 
{\bf A}, 226. 
Lifhsitz, I.M., 1962, {\it Zh. Eksp. Teor. Fiz.} {\bf 42}, 1354 
({\it Sov. Phys.} JETP {\bf 15}, 939). \\
Metropolis N., Rosenbluth A.W., Rosenbluth M.N., Teller, A.H and Teller E., 
1953, {\it J. Chem. Phys.} {\bf 21}, 1087.\\
Potts R.B., 1952, {\it Proc. Camb. Phil. Soc.} {\bf 48}, 106.\\
Runnels, L.K., 1983 {\it Phase transitions and critical phenomena} vol. 8,
ed C. Domb and J.L. Lebowitz (New York, Academic) p. 305.\\
Wang J.S., Nielaba P. and Privman V., 1993, {\it Physica} {\bf A 199},
527.\\
Wang J.S., Nielaba P. and Privman V., 1993, {\it Mod. Phys. Lett.} 
{\bf B 7}, 189.\\
Wu F.Y., 1982, {\it Rev. Mod. Phys.} 54, 235.\\
Yeomans J.M., 1992, {\it Statistical Mechanics of Phase Transitions} 
(New York: Oxford Un. Press).
 

\newpage

\begin{figure}
\centerline{\psfig{file=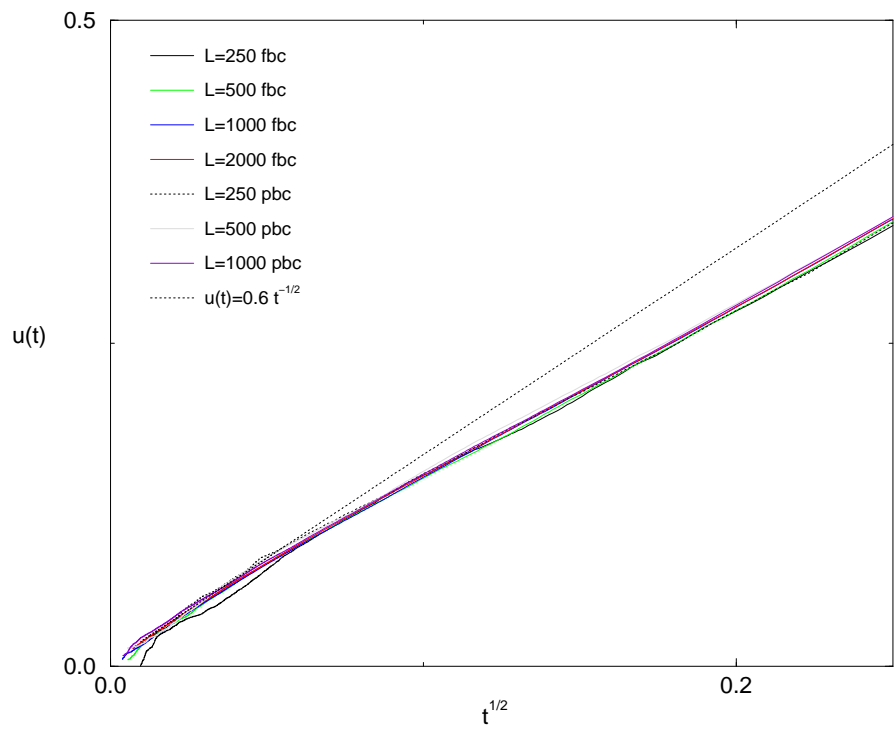,height=14cm,angle=-90}}
\caption{Behaviour of the energy per site with time, for a $q=3$ 
Potts model on a square lattice.
Different curves correspond to different size ($L$) and boundary
conditions ($pbc=$ periodic, $fbc=$ fixed). The dashed line is an extrapolation
to $t=\infty$ of the $1/\sqrt(t)$ regime at large times.}
\label{fig1}
\end{figure}

\begin{figure}
\centerline{\psfig{file=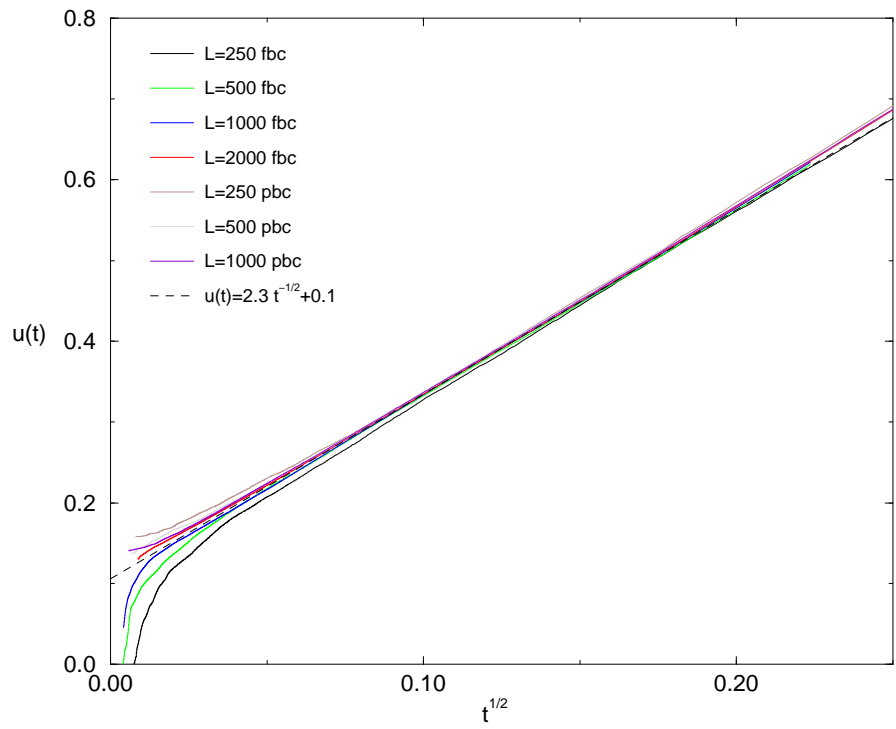,height=14cm,angle=-90}}
\caption{The same as figure 1 for q=7.}
\label{fig2}
\end{figure}

\begin{figure}
\centerline{\psfig{file=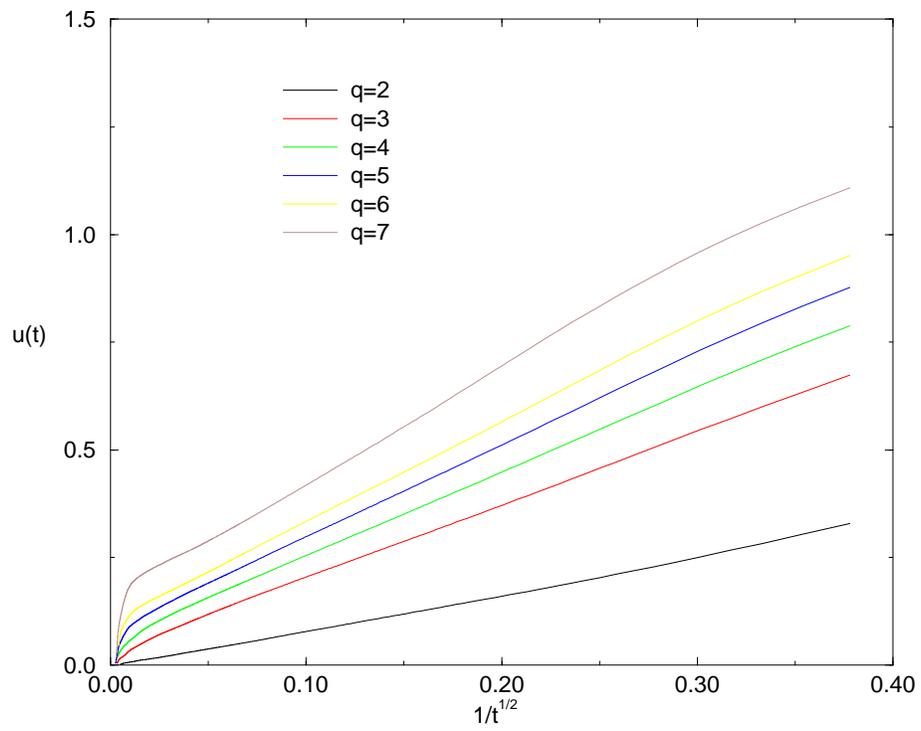,height=14cm,angle=-90}}
\caption{The increase of the excess energy 
for increasing $q$, when $t \rightarrow \infty$, is
well visible from the time behaviour of the energy per site 
for fixed boundary conditionss ($L=1000$).}
\label{fig3}
\end{figure}

\end{document}